\documentclass[12pt]{article}

\usepackage{epsfig}
\include{amssym}

\def\mathbf{\vec}

\def\ca{\c{c}\~{a}}

\newcommand{\st}{\mathrm{st}}

\newcommand{\eff}{\mathrm{eff}}
\newcommand{\ud}{\mathrm{d}}

\begin{document}

\centerline{\Large\bf Stationary phase corrections in the process }
\vspace{0.2cm}
\centerline{\Large\bf of bosonization of multi-quark interactions}
\vspace{1cm}

\centerline{\large 
            A. A. Osipov$^\dagger$\footnote{On leave from Joint 
            Institute for Nuclear Research, Laboratory of Nuclear 
            Problems, \mbox{\ \ \ \ }  
            141980 Dubna, Moscow Region, Russia. 
            Email address: osipov@nusun.jinr.ru},
            B. Hiller$^\dagger$\footnote{Email address: 
            brigitte@teor.fis.uc.pt}, 
            J. Moreira$^\dagger$\footnote{
            Email address:jmoreira@teor.fis.uc.pt},
            A. H. Blin$^\dagger$\footnote{Email address:
            alex@teor.fis.uc.pt}} 
\vspace{0.5cm}
\centerline{$^\dagger$\small\it Centro de F\'{\i}sica Te\'{o}rica, 
            Departamento de
            F\'{\i}sica da Universidade de Coimbra,}
\centerline{\small\it 3004-516 Coimbra, Portugal}
\vspace{1.5cm}

\centerline{\bf Abstract}
\vspace{0.5cm}

The functional integration over the auxiliary bosonic variables of
cubic order related with the effective action of the  Nambu -- 
Jona-Lasinio model with 't Hooft term has recently been obtained in
the form of a loop expansion. Even numbers of loops contribute
to the action, while odd numbers of loops are assigned to the measure. 
We consider the two-loop corrections and analyse their effect on the 
low-lying pseudoscalar and scalar mass spectra, quark condensates and 
weak decay constants. The results are compared to the leading order 
calculations and other approaches.
 
\vspace{4.0cm}
\noindent
PACS number(s): 12.39.Fe, 11.30.Rd, 11.30.Qc

\newpage 

\section*{\centerline{\large\sc 1. Introduction}}

The large distance dynamics of QCD is dictated to a great extent by
the spontaneous symmetry breaking of chiral symmetry 
\cite{Weinberg:1979,Gasser:1983}. The Nambu -- Jona-Lasinio (NJL)
model of fermionic fields \cite{Nambu:1961} suggests that the dynamical 
mechanism for such breaking be in analogy with the Ginsburg -- Landau
theory of superconductivity \cite{Ginsburg:1950}. Numerous studies 
\cite{Eguchi:1976}-\cite{Ebert:1986} have been performed since that
time with the respective effective mesonic action derived from
four-quark interactions of the NJL type. During these years the
resolution of $U_A(1)$ problem has been found and, in particular, 
the relevance of the $U(3)_L\times U(3)_R$ chiral symmetric NJL model 
combined with the six-quark 't Hooft flavour determinantal interaction 
(NJLH) \cite{Hooft:1978} for low-energy phenomenology of mesons was noted 
\cite{Bernard:1988}-\cite{Osipov:2004NPA}. The explicit breaking of
the unwanted $U_A(1)$ axial symmetry by the 't Hooft determinant is 
motivated by the instanton approach to low-energy QCD
\cite{Hooft:1978}, \cite{Diakonov:1985}.      

Originally written in terms of fermionic degrees of freedom, the NJLH 
model has been widely explored at mean field level with Bethe-Salpeter 
and Hartree-Fock techniques applied to quark-antiquark scattering in
its various channels of interaction \cite{Bernard:1988}, 
\cite{Klimt:1990,Bernard:1993}. 

In parallel, functional integral methods have been used to obtain the 
Lagrangian in bosonized form \cite{Reinhardt:1988}, 
\cite{Osipov:2002}-\cite{Osipov:2005a}. The bosonization gives rise to 
a doubling of the mesonic auxiliary fields, of which one set has to be 
integrated out. This latter, in the presence of the 't Hooft
interaction, involves a term of cubic order, which cannot be
integrated out exactly. In \cite{Reinhardt:1988} the leading order 
stationary phase approximation (SPA) was calculated. At this order the 
effective potentials obtained with both methods coincide 
\cite{Osipov:2004}.

Given its success in describing a large bulk of empirical data, the 
question arises whether corrections to the leading order SPA result are
small. By embarking in this task, a series of startling facts came
across our investigations \cite{Osipov:2005a}:

(i) The stationary phase equations which one obtains in the NJLH model 
have more than one root (critical point). Only one has a regular
behaviour in the limit $\kappa\to 0$ of the six-quark coupling, the 
others are singular. The rigorous SPA treatment requires taking into 
account all critical points, which give rise to an unstable vacuum for 
the theory.
 
(ii) The result obtained in \cite{Reinhardt:1988} corresponds to the
regular root contribution. It is an approximation which leads to an
effective potential with a well separated local minimum, which 
approaches smoothly the stable NJL vacuum as $\kappa\to 0$. Such a local
minimum is probably a good ground for phenomenological estimates, 
at least all known calculations made in the NJLH model are based on this 
approximation. It has been shown recently \cite{Osipov:2005} that  
eight-quark interactions stabilize this vacuum state, opening the way 
to justify this approach theoretically.
 
(iii) Two expansions of the effective action have been considered: the 
perturbative series in $\kappa$ and the loop expansion. Both of them
were never studied beyond the leading order. It is tacitly assumed
that next to the leading order corrections are small, although this
fact has never been proven.  

In this paper we quantify the two-loop order contributions to 
the Lagrangian derived previously by studying their impact on the 
mass spectrum of low-lying mesons. We show that the effect is of the 
order of a few percent compared to the leading order masses, improving 
them slightly. We think it is rather safe to conclude that the loop
expansion is rapidly converging, at least for the mass spectra. 

The paper is structured as follows. In the next section we collect the 
essential information needed to extract the linear and quadratic terms 
which contribute to the gap equations and mass terms respectively. In 
section 3 we write out the expressions for the gap equations, masses, 
weak decay constants and condensates. Section 4 contains the numerical 
results and discussion. Conclusions are presented in section 5.


\section*{\centerline{\large\sc 2. The "tandem" Lagrangian}}

To be self-contained and define the notation we review the main
ingredients of our model calculations. The underlying multi-quark 
Lagrangian is bosonized in a two step (tandem) process, in which a 
semi-bosonized functional, quadratic in the fermionic fields,  and
another functional, depending only on the auxiliary bosonic variables, 
can be dealt with separately. The integration over the quadratic
fermionic degrees of freedom is formally exact and is calculated using 
a generalized heat kernel method. We start by presenting how these two 
types of functionals emerge and how we obtain and calculate the loop 
expansion we are after.    


\subsection*{\normalsize\it 2.1 The stationary phase contribution}

We consider the fermionic Lagrangian  
\begin{equation}
\label{totlag}
  {\cal L}_{NJLH}=\bar{q}(i\gamma^\mu\partial_\mu -m)q
          +{\cal L}_{NJL}+{\cal L}_H,
\end{equation}
which contains the NJL four-quark vertices of the scalar and 
pseudoscalar types
\begin{equation}
\label{L4q}
  {\cal L}_{NJL}=\frac{G}{2}\left[(\bar{q}\lambda_aq)^2+
                           (\bar{q}i\gamma_5\lambda_aq)^2\right]
\end{equation}
and the six-quark 't Hooft interaction \cite{Hooft:1978}
\begin{equation}
\label{Ldet}
  {\cal L}_{H}=\kappa (\mbox{det}\ \bar{q}P_Lq
                         +\mbox{det}\ \bar{q}P_Rq),
\end{equation}
where $m$ is the diagonal current quark matrix for quark fields 
with $N_f=3$ flavours and  $N_c=3$ colours. In eq.(\ref{L4q})
$\lambda_a$, $a=0,1...8$, are the normalized ($\mbox{tr} \lambda_a
\lambda_b=2\delta_{ab}$) matrices in flavour space. The explicit form
of these $U(3)$ hermitian generators is $\lambda_0=\sqrt{2/3}$ and 
$\lambda_a$ for $a\neq 0$ are the usual Gell-Mann matrices.  
The positive coupling $G, [G]=\mbox{GeV}^{-2}$ has order $G\sim 1/N_c$. In 
(\ref{Ldet}) the negative coupling $\kappa$ of dimension $[\kappa]=
\mbox{GeV}^{-5}$ has the large $N_c$ asymptotics $\kappa\sim
1/N_c^{N_f}$. Therefore ${\cal L}_{NJL}$ dominates over ${\cal L}_H$
at large $N_c$. The matrices $P_{L,R}=(1\mp\gamma_5)/2$ are projectors 
on the chiral states and the determinant is over flavour indices.  

We are assuming that the quark fields transform like the fundamental 
representations of the global $U(3)_L\times U(3)_R$ chiral group,
i.e.,
\begin{equation}
\label{chtr-q}
   \delta q=i(\alpha +\gamma_5\beta )q, \qquad
   \delta \bar{q} =-i\bar{q}(\alpha -\gamma_5\beta ),   
\end{equation}
where the parameters of the infinitesimal transformations are chosen as
$\alpha =\alpha_a\lambda_a$, $\beta =\beta_a\lambda_a$. One now
observes that
\begin{equation}
   \delta {\cal L}_{NJLH} = i\bar q\left([\alpha ,m]-\gamma_5\{\beta
   ,m\} \right)q + 2i\sqrt{6}\beta_0\kappa\left(
   \mbox{det}\ \bar{q}P_Rq
                         -\mbox{det}\ \bar{q}P_Lq\right).
\end{equation}
The global chiral symmetry is broken explicitly by the current quark
mass term and the $U(1)_A$ axial symmetry is broken too due to the 
't Hooft interaction.

The functional integral in bosonized form is derived in 
\cite{Reinhardt:1988}, and has the form
\begin{eqnarray}
\label{genf3}
   Z&\!\!\! =\!\!\!&\int \prod_A{\cal D}\Pi_A 
                    {\cal D}q{\cal D}\bar{q}\
                    \exp\left(i\int\ud^4x
                    {\cal L}_q(\bar{q},q,\sigma ,\phi )\right)
                    \nonumber \\
    &\!\!\!\times\!\!\!&
    \int\limits^{+\infty}_{-\infty}\prod_A{\cal D}R_A\
     \exp\left(i\int\ud^4x{\cal L}_r(\Pi ,\Delta ;R)\right),
\end{eqnarray}  
where
\begin{eqnarray}
\label{lagr2}
  {\cal L}_q &\!\!\! =\!\!\!&
  \bar{q}(i\gamma^\mu\partial_\mu -M-\sigma - i\gamma_5\phi )q, \\
\label{Lr}
   {\cal L}_r &\!\!\! =\!\!\!& R_A (\Pi_A+\Delta_A) + \frac{G}{2}\ R_A^2 
   + \frac{\kappa}{3!}\ \Phi_{ABC}R_AR_BR_C.
\end{eqnarray}
with a cubic polynomial in the fields $R_A$ in the exponent. The
notation is as follows \cite{Osipov:2005a}: $R_A=(R_a,R_{\dot{a}})=
(s_a,p_a)$ and $\Pi_A=(\Pi_a,\Pi_{\dot{a}})=(\sigma_a,\phi_a)$ are a 
very compact way to represent two sets of auxiliary bosonic variables, 
each containing a scalar $s_a$ ($\sigma_a$) and a pseudoscalar $p_a$ 
($\phi_a$) nonet. The indices $(a, \dot{a})$ run  from $0$ to $8$ 
independently in flavour space. We also define the related quantity
$\Delta_A=(\Delta_a,0)=(M_a- m_a,0)$. 

The external scalar fields $\sigma =\sigma_a\lambda_a$ have been 
shifted $\sigma_a\rightarrow\sigma_a +M_a$ by the constituent quark 
mass $M_a$, so that the expectation value of the shifted fields in the 
vacuum corresponding to dynamically broken chiral symmetry vanish. The 
vacuum expectation value of the ``unshifted'' scalar field 
\begin{equation}
\label{vev}
   \big <\sigma \big >=M_a\lambda_a=\mbox{diag}(M_u,M_d,M_s) 
\end{equation}
gives the point where the effective potential of the model $V(\big <
\sigma\big >)$ achieves its local minimum. The corresponding condition 
is known as the ``gap'' equation. It eliminates tadpole graphs and 
determines the values of constituent quark masses as functions of the
model parameters and of the cutoff $\Lambda$. The case $m_u\neq m_d\neq
m_s$ corresponds to the most general breakdown of the $SU(3)$ flavour 
symmetry, giving $M_u\neq M_d\neq M_s$. In this way the ground state of
the system includes effects of the explicit symmetry breaking. We will
assume in the following that $m_u=m_d$.       

The variables $\sigma_a$ and $\phi_a$ must be replaced by the physical 
scalar and pseudoscalar states $\sigma_a^{ph}$ and $\phi_a^{ph}$ 
determined through the appropriate normalization of their kinetic terms. 
Note that these terms, as well as other important contributions to the meson 
masses and interactions of the effective mesonic Lagrangian, are obtained as a result of 
integration over the quark fields in $Z$. One has 
\begin{equation}
\label{ren}
   \sigma_a,\ \phi_a = g \sigma_a^{ph},\ g\phi_a^{ph}
\end{equation}
at leading order of the heat kernel 
expansion of the effective mesonic action (see next section). The 
quark-meson coupling $g$, being a function of parameters of the model, 
fulfills in addition the Goldberger -- Treiman relation at the quark 
level: $g=M_u/f_\pi$. Combining this relation with (\ref{vev}) and 
(\ref{ren}) one finds the well-known linear sigma model result 
\cite{Geffen:1969}
\begin{equation}
\label{lsm}
   \big <\sigma_u^{ph} \big >=f_\pi\, . 
\end{equation}

Finally, the three index coefficients $\Phi_{ABC}$ are defined as  
\begin{equation} 
\label{Phi}
   \Phi_{abc} = -\Phi_{a\dot{b}\dot{c}}=\frac{3}{16}A_{abc}\, , \quad 
   \Phi_{ab\dot{c}}= \Phi_{\dot{a}\dot{b}\dot{c}}=0, 
\end{equation}
obeying
\begin{equation}
\label{contr}
      \Phi_{ABC}\delta_{BC}=0.
\end{equation}
The totally symmetric constants $A_{abc}$ are related to the flavour 
determinant, and equal to
\begin{equation}
\label{A}
   A_{abc}=\frac{1}{3!}\ \epsilon_{ijk}\epsilon_{mnl}(\lambda_a)_{im}
             (\lambda_b)_{jn}(\lambda_c)_{kl}. 
\end{equation}

Now the functional integral over the auxiliary variables $R_A$ in 
(\ref{genf3}) 
\begin{equation}
\label{intJi}
     {\cal Z}[\Pi ,\Delta ]\equiv 
     \int\limits^{+\infty}_{-\infty}\prod_A{\cal D}R_A\
     \exp\left(i\int\ud^4x{\cal L}_r(\Pi ,\Delta ;R)\right),
\end{equation} 
can be written in the form \cite{Osipov:2005a}
\begin{eqnarray}
\label{Zloop1}
     {\cal Z}[\Pi ,\Delta ]&\!\!\!\sim \!\!\!&  
     \exp\left( i\int\ud^4x {\cal L}_\st\right)
     \nonumber \\
     &\!\!\!\times\!\!\!&
     \int\limits^{+\infty}_{-\infty}\prod_A{\cal D}\bar{R}_A\
     \exp\left(\frac{i}{2}\int\ud^4x{\cal L}_{AB}''
     \bar{R}_A \bar{R}_B \right) \nonumber \\
     &\!\!\!\times\!\!\!& \sum_{n=0}^\infty \frac{1}{n!}
     \left(i\frac{\kappa}{3!}\ \Phi_{ABC}\int\ud^4x 
     \bar{R}_A\bar{R}_B\bar{R}_C\right)^n.  
\end{eqnarray}
Here  ${\cal L}_\st$ is the stationary value of the Lagrangian 
${\cal L}_r(\Pi ,\Delta ;R)$ associated with the regular critical
point, around which the effective Lagrangian has been expanded. The 
barred fields indicate that they are shifted with respect to their 
stationary values $\bar R =R-R_{\st}$. We denote by ${\cal L}_{AB}''$
the coefficient of the expansion to second order in the fields. The 
expansion stops exactly at the third order ${\cal L}_{ABC}'''=\kappa
\Phi_{ABC}/3!$, whose exponent is represented here as an infinite
series. 

The terms with odd values of $n$ do not contribute in
(\ref{Zloop1}), because the corresponding functional integrals over
$\bar{R}_A$ equal to zero. The first term, $n=0$, sums all tree  
diagrams of a perturbative series in powers of the coupling $\kappa$
resulting in ${\cal L}_\st$. The $n=2$ term represents 
the first non-leading correction to the effective Lagrangian (for 
details we refer to \cite{Osipov:2005a}, where we show that this
correction can be associated with a ``two-loop'' contribution of 
quantum auxiliary bosonic fields)
\begin{equation}
\label{2loop}
   {\cal L}_\eff = {\cal L}_\st + \left(\frac{\lambda}{2\pi}\right)^8 
   \frac{3\kappa^2 {\cal M}}{32N(N+2)(N+4)}\ . 
\end{equation}

The stationary Lagrangian reads to cubic order in the fields 
\cite{Osipov:2002,Osipov:2004} 
\begin{eqnarray}
\label{lam}
   {\cal L}_{\st}
   &=&h_a\sigma_a
      +\frac{1}{2}h_{ab}^{(1)}\sigma_a\sigma_b  
      +\frac{1}{2}h_{ab}^{(2)}\phi_a\phi_b \nonumber \\
   &+&\frac{1}{3}\sigma_a\left[h^{(1)}_{abc}\sigma_b\sigma_c
      +\left(h^{(2)}_{abc}+h^{(3)}_{bca}\right)\phi_b\phi_c\right]
      \nonumber \\
   &+&{\cal O}(\mbox{field}^4).
\end{eqnarray}  

The two-loop corrections, contained in the second term of
(\ref{2loop}), consist of a rather intricate dependence
\begin{equation}
\label{calm}
   {\cal M} =  
   \left(\mbox{tr}\left.{\cal L}''\right.^{-1}\right)^3\!
   + 6\,\mbox{tr}\left.{\cal L}''\right.^{-1}
   \mbox{tr}(\left.{\cal L}''\right.^{-1})^2\!      
   + 8\,\mbox{tr}\left(\left.{\cal L}''\right.^{-1}\right)^3
\end{equation}
on flavour traces of powers of $\left.{\cal L}''\right.^{-1}$, which 
is the inverse of the real and symmetric $N\times N$ matrix  ($N=18$)
\begin{equation}
\label{Qab}
  {\cal L}''_{AB}(R_{\st})=
  \left(
  \begin{array}{cc}
  G\delta_{ab}+\displaystyle\frac{3\kappa}{16}A_{abc}s_{\st}^{c}
  &-\displaystyle\frac{3\kappa}{16}A_{abc}p_{\st}^{c}\\
   -\displaystyle\frac{3\kappa}{16}A_{abc}p_{\st}^{c}
  &G\delta_{ab}-\displaystyle\frac{3\kappa}{16}A_{abc}s_{\st}^{c}
  \end{array}
  \right),
\end{equation} 
calculated at the stationary points $s^a_{\st}$ and $p^a_{\st}$. These 
are expressed in increasing powers of the external fields $\sigma_a , 
\phi_a$  
\begin{eqnarray}
\label{rst}
   s^a_{\st}
    &=&h_a+h_{ab}^{(1)}\sigma_b
       +h_{abc}^{(1)}\sigma_b\sigma_c
       +h_{abc}^{(2)}\phi_b\phi_c 
       \nonumber \\
    &+&h_{abcd}^{(1)}\sigma_b\sigma_c\sigma_d
       +h_{abcd}^{(2)}\sigma_b\phi_c\phi_d
       +\ldots \\
   p^a_{\st}&=&h_{ab}^{(2)}\phi_b
       +h_{abc}^{(3)}\phi_b\sigma_c
       +h_{abcd}^{(3)}\sigma_b\sigma_c\phi_d
       \nonumber \\
     &+&h_{abcd}^{(4)}\phi_b\phi_c\phi_d
       +\ldots 
\end{eqnarray}
with $h^{(i)}_{ab\ldots}$ depending on $\Delta_a$ and coupling
constants (see Appendix). In particular the coefficients have 
nonvanishing components for $a=(0,3,8)$ and are obtained, with 
$h=h_a\lambda_a=\mbox{diag}(h_u,h_d,h_s)$, in the case of isotopic
symmetry ($h_u=h_d$) as \cite{Osipov:2004} 
\begin{equation}
\label{saddle-1}
  \left\{
         \begin{array}{rcl}
        && Gh_u+\Delta_u+\displaystyle\frac{\kappa}{16}\ h_uh_s=0, \\
\\        
        && Gh_s+\Delta_s+\displaystyle\frac{\kappa}{16}\ h_u^2=0. \\
         \end{array}
  \right. 
\end{equation}

In eq.(\ref{2loop}) $\lambda$ denotes an ultra-violet cutoff associated 
with the stationary phase corrections to the functional integral over 
auxiliary bosonic fields. It is a free parameter to be fixed by 
phenomenology.

To handle the new contribution ${\cal M}$, we expand  
${\cal L}''_{AB}(R_{\st})$, which we abbreviate from now on as 
${\cal L}''$, to second order in the external fields $\sigma_a,\phi_a$, 
\begin{equation}
\label{expandL}
   {\cal L}''=L_0+L_1+L_2+{\cal O}(\mbox{field}^3),
\end{equation}
and its inverse 
\begin{equation}
\label{expandLinv}
   {\cal L}''^{-1}=\bar {L}_0+\bar {L}_1+\bar {L}_2
                  +{\cal O}(\mbox{field}^3).
\end{equation}
This is all one needs to extract the relevant terms to the masses 
arising from the two-loop correction term. Here $L_i$, ($i=0,1,2$), 
denote the matrices which are constant, linear and quadratic in the 
fields respectively. The  $\bar {L}_i$ are constructed order by order, 
starting from the $0$th order 
\begin{equation}
\label{0ord}
{\cal L}''{\cal L}''^{-1}=L_0 \bar {L}_0 =1,
\end{equation}
i.e. $\bar {L}_0=L_0^{-1}$. 

The next terms are conditioned by this relation. Combining the first 
order Lagrangians and truncating at the 
linear fields 
\begin{equation}
\label{1ord}
   (L_0+L_1)(\bar {L}_0+\bar {L}_1)\rightarrow L_0 \bar {L}_0 + L_1 
   \bar {L}_0 +L_0 \bar {L}_1=1
\end{equation}
one gets the matrix $\bar {L}_1$, after using ($\ref{0ord}$),
\begin{equation}
\label{invL1}
\bar {L}_1=-\bar {L}_0 L_1 \bar {L}_0. 
\end{equation}

In a similar fashion one derives the  matrix $\bar {L}_2$ as
\begin{equation}
\label{invL2}
   \bar {L}_2=-\left(\bar {L}_0 L_2\bar {L}_0 +\bar {L}_0 L_1 
   \bar {L}_1\right)=-\left(\bar {L}_0 L_2\bar {L}_0-\bar {L}_0 
   L_1\bar {L}_0 L_1\bar {L}_0\right).
\end{equation}

Using $\bar {L}_i$ in (\ref{expandLinv}) and inserting in
(\ref{calm}), one obtains
\begin{eqnarray}
\label{M2}
   {\cal M}_2 &\!\!\! =\!\!\!& 3\left\{
     \left(\mbox{tr}\bar{L}_0\right)\!\left(\mbox{tr}\bar{L}_1\right)^2
     +\left(\mbox{tr}\bar{L}_0\right)^2\!\left(\mbox{tr}\bar{L}_2\right) 
     + 8\ \mbox{tr}\left(\bar {L}_0{\bar {L}_1}^2 
     + \bar{L}_2{\bar{L}_0}^2\right)\right\} \nonumber \\
  &\!\!\! +\!\!\!& 6 \left\{ \mbox{tr}\left(\bar{L}_0\right) 
     \!\left[\mbox{tr}\left(\bar {L}_1^2\right)
     + 2\ \mbox{tr}\left(\bar{L}_0\bar{L}_2\right)\right]
     \right.\nonumber \\
  &\!\!\! +\!\!\!& \left. 2\ \mbox{tr}\left(\bar{L}_1\right)
     \mbox{tr}\left(\bar{L}_0\bar{L}_1\right)
     +\left(\mbox{tr}\bar{L}_2\right)
     \mbox{tr}\left(\bar{L}_0^2\right)\right\},
\end{eqnarray}
where ${\cal M}_2$ stands for the part of ${\cal M}$ which
contains only the second order terms in the fields $\sigma_a,\phi_a$.

This expression is used in ``Mathematica" \cite{Mathematica}. Although 
the results, after evaluating of traces, are analytical they are very 
lengthy and not illuminating, and will not be presented here. However 
some structures are relevant for the low-energy theorems and the
results will be encrypted in them, as shown in the section where we 
present the mass formulae.


\section*{\normalsize\it 2.2  The heat kernel contribution }

It still remains to evaluate the functional integral over the quark 
degrees of freedom in eq.(\ref{genf3}). The Lagrangian ${\cal L}_q$ 
is invariant under the chiral transformations (\ref{chtr-q}) and the
transformations 
\begin{equation}
\label{trsp}
   \delta\sigma =i[\alpha ,\sigma +M] + \{\beta ,\phi\}, \qquad
   \delta\phi   =i[\alpha ,\phi ] - \{\beta ,\sigma +M\},
\end{equation}
induced by them for the external fields. All symmetry breaking terms
have been absorbed in ${\cal L}_r$. This fact is of importance, since
one can use then the generalized asymptotic expansion of the quark 
determinant \cite{Osipov1:2001,Osipov2:2001}. This method preserves the
above mentioned symmetry at any order, taking into account the effects 
of the flavour symmetry breaking contained in the mass matrix $M$.   
Thus the corresponding part of the effective action can be written 
\begin{equation}
\label{ZY}
   \ln|\det D|=-\frac{1}{32\pi^2}\int\ud^4 x_E\sum_{i=0}^{\infty}
        I_{i-1}\mbox{tr}(b_i),
\end{equation}
where $D=i\gamma_\mu\partial_\mu -M-\sigma -i\gamma_5\phi$ is the
Dirac operator present in ${\cal L}_q$, eq.(\ref{lagr2}), and the
$b_i$ are generalized Seeley-DeWitt coefficients \cite{Osipov1:2001}, 
of which we show the first four for the case of $SU(2)_I\times U(1)_Y$ 
flavour symmetry  
\begin{eqnarray}
\label{SW}
   b_0\!\!\!\!\!\!\!\!
      &&=1,  \nonumber\\
   b_1\!\!\!\!\!\!\!\!
      &&=-Y, \nonumber\\
   b_2\!\!\!\!\!\!\!\!
      &&=\frac{Y^2}{2}+\frac{\Delta_{us}}{\sqrt{3}}\ \lambda_8 Y,
         \nonumber\\
   b_3\!\!\!\!\!\!\!\!
      &&=-\frac{Y^3}{3!}+\frac{\Delta_{us}^2}{6\sqrt{3}}\
      \lambda_8 Y-\frac{\Delta_{us}}{2\sqrt{3}}\ \lambda_8 Y^2
      -\frac{1}{12}\left(\partial Y\right)^2.
\end{eqnarray}
In the present case the background dependent structure $Y$ is given by 
\begin{equation}
   Y=i\gamma_{\mu}(\partial_{\mu} \sigma+i\gamma_5\partial_{\mu}\phi)
    +\sigma^2+\{M,\sigma\}+\phi^2+i\gamma_5[\sigma+M,\phi ].
\end{equation}
We use the definition $\Delta_{ij}\equiv M_i^2-M_j^2$. In
eq.(\ref{ZY}) the trace is to be taken over colour, flavour and Dirac 
4-spinor indices and the regulator-dependent integrals $I_i$ are 
the weighted sums 
\begin{equation}
\label{Iint}
   I_i=\frac{1}{3}\left(2J_i(M_u^2)+J_i(M_s^2)\right)
\end{equation}
with 
\begin{equation}
\label{Ji}
   J_i(M_j^2)=\int\limits_{0}^{\infty}\frac{\ud t}{t^{2-i}}\rho (t\Lambda^2)
              \exp (-tM_j^2).
\end{equation}
They are regularized with the Pauli-Villars regularization scheme 
\cite{Pauli:1949} with two subtractions and one ultra-violet cutoff 
$\Lambda$   
\begin{equation}
   \rho (t\Lambda^2)=1-(1+t\Lambda^2)\mbox{exp}(-t\Lambda^2).
\end{equation}

We obtain, for instance, \cite{Osipov:2000a}
\begin{eqnarray}
\label{J0}
   J_0(M^2)\!\!\!\!\!\!\!\!
     &&=\Lambda^2-M^2\ln\left(1+\frac{\Lambda^2}{M^2}\right), \\
   J_1(M^2)\!\!\!\!\!\!\!\!
     &&=\ln\left(1+\frac{\Lambda^2}{M^2}\right)
             -\frac{\Lambda^2}{\Lambda^2+M^2}\ .
\end{eqnarray}
Both of them are divergent in the limiting case $\Lambda\to\infty$. 
Note that $\Lambda$ does not need to be the same cutoff as $\lambda$ 
of eq.(\ref{2loop}). In the following we restrict our study to the 
two nontrivial terms, $b_1$ and $b_2$, in the asymptotic expansion of 
$\ln|\det D|$. In this case only $I_0$ and $I_1$ are involved, related 
to the quark one-loop integrals of one- and two-point functions 
respectively, at zero four-momentum transfer.


\section*{\centerline{\large\sc 3. Gap equations, 
          condensates, meson spectra}}



\section*{\normalsize\it 3.1 Gap equations and condensates}

The complete effective bosonized Lagrangian  
\begin{equation}
\label{lg}
{\cal L}_b={\cal L}_{HK}+{\cal L}_\st+{\cal L}_c
\end{equation}
comprises contributions from the heat kernel expansion to order $b_2$, 
${\cal L}_{HK}$,  and from (\ref{2loop}), where ${\cal L}_c$ stands
for the two-loop corrections. We restrict to the case $SU(2)_I\times
U(1)_Y$ symmetry, e.g. $M_u=M_d\ne M_s$. The first two contributions 
remain the same as in the leading order calculations 
\cite{Osipov:2004NPA}. 

Equating the coefficient of $\sigma_i$, $i=(u,d,s)$ in eq.(\ref{lg}) 
to zero we obtain the gap equations
\begin{eqnarray}
\label{gapfluc}
   &&h_u +\frac{N_c}{6\pi^2}\ M_u\left[3I_0+(M_s^2-M_u^2)I_1\right]
         + 2 c_u = 0, \nonumber \\
   &&h_s +\frac{N_c}{6\pi^2}\ M_s\left[3I_0-2(M_s^2-M_u^2)I_1\right]
         + 2 c_s = 0,
\end{eqnarray}
where $c_u$ and $c_s$ denote the corrections arising from ${\cal  L}_c$. 
They depend on $h_u, h_s, \lambda, \kappa$. These equations must be 
solved self-consistently and in conjunction with the stationary phase 
conditions (\ref{saddle-1}). The solutions $M_i$ of (\ref{gapfluc}) 
allow to calculate the condensates $\big <\bar {u} u\big >$ and 
$\big <\bar {s} s\big >$ (see eq.(\ref{conf}))
\begin{equation}
\label{cond}
   \big <\bar {q}_i q_i\big > = -\frac{N_c}{4\pi^2}\left[
   M_i J_0(M_i^2)-m_i J_0(m^2_i)\right], 
\end{equation}
where we have subracted the contribution from the trivial vacuum 
\cite{Bernard:1988}. Although they are structurally identical to the 
condensates calculated at leading order, they encode the information
of the correction terms $c_i$ implicitly through $M_i$.


\section*{\normalsize\it 3.2 Meson masses}

The expressions for the leading order masses, i.e. with ${\cal L}_c$
put to zero, will not be repeated here. They were obtained in 
\cite{Osipov:2004NPA}. The correction mass terms can just be added to
the leading order terms in their "raw" form, that is, as they are 
directly extracted from ${\cal L}_{HK}$, depending on $I_0,I_1$ 
integrals. To check the low energy theorems, one can then use the new 
gap equations, with the correction terms $c_i$ included, to eliminate 
these integrals. For example for the pion, $\phi_j\ (j=1,2,3)$, one 
has
\begin{equation}
\label{mpihk}
   {\cal L}_{HK}(m^2_\pi)=\frac{N_c}{12\pi^2}
   \left(3I_0+\Delta_{su} I_1\right) \phi_j^2, 
\end{equation}
\begin{equation}
\label{mpist}
{\cal L}_{\st}(m^2_\pi)= -\frac{\phi_j^2}{2 G (1+\omega_s)}\ .
\end{equation}

With ``Mathematica" we are able to identify 
\begin{equation}
\label{mpic}
   {\cal L}_{c}(m^2_\pi)=-\frac{c_u \phi_j^2}{(4G)^2\omega_u(1+\omega_s)}\ 
   ,
\end{equation}
where $\omega_i=\kappa h_i/(16 G)$. 
This connection between the pion mass correction to the gap equation 
correction term $c_u$ is crucial to guarantee the Goldstone limit.
Indeed we obtain, after eliminating $I_0,I_1$ from (\ref{mpihk}) with 
help of the gap equations,
\begin{eqnarray}
\label{mpi2}
   m^2_{\pi}&=&{\stackrel{o} m}^2_{\pi}
   \left(1+2\ \frac{c_u}{h_u}\right),\nonumber \\
   {\stackrel{o} m}^2_{\pi}&=&\frac{g^2m_u}{M_u G(1+\omega_s)}\ ,
\end{eqnarray}
where ${\stackrel{o} m}_{\pi}$ is structurally identical with the
leading order pion mass; $g^2=4\pi^2/(N_c I_1)$ renormalizes the pion 
fields to the physical fields (see eq.(\ref{ren})). We used also that 
\begin{equation}
   -\frac{h_u}{\Delta_u}=\frac{1}{G(1+\omega_s)}
\end{equation}
which is a simple consequence of the stationary phase conditions 
(\ref{saddle-1}). 

In an analogous way we are able to get for the kaon mass
\begin{eqnarray}
\label{mk2}
   m^2_{K}&\!\!\! =\!\!\!&{\stackrel{o} m}^2_K \left(1+2\ 
             \frac{c_u+c_s}{h_u+h_s}\right), \nonumber \\
   {\stackrel{o} m}^2_K&\!\!\! =\!\!\!&
   \frac{g^2(m_u+m_s)}{G(M_u+M_s)(1+\omega_u)}\ ,
\end{eqnarray}
where ${\stackrel{o} m}^2_K$ has the form of the leading order kaon
mass. We used again (\ref{saddle-1}) to obtain
\begin{equation}
   -\frac{h_u+h_s}{\Delta_u+\Delta_s}=\frac{1}{G(1+\omega_u)}\ .
\end{equation}

Concerning the $\eta,\eta'$ corrections, we show here only some
relevant properties obtained with the help of ``Mathematica" in the 
SU(3) limit
\begin{eqnarray}
\label{etas}
   &&(\Delta m^2_\pi)_c=(\Delta m^2_K)_c=(\Delta m^2_{88})_c,
     \nonumber \\
   &&(\Delta m^2_{08})_c=0,
     \nonumber \\
   &&(\Delta m^2_{00})_c-(\Delta m^2_{88})_c\ne 0,
\end{eqnarray}
where, for instance, $(\Delta m^2_\pi )_c$ is the contribution to the
pion mass obtained from the Lagrangian ${\cal L}_c$.
Therefore the correction to the flavour $(0,8)$ components follow the 
same patterns as in the leading order case \cite{Osipov:2004NPA}, 
complying with the low-energy requirements: the octet member
$m^2_{88}$ remains degenerate with the pion and kaon, the mixing 
$m^2_{08}$ vanishes and in the chiral limit the correction to the mass
of the singlet $m^2_{00}$ is also non-vanishing, and will therefore 
contribute to the singlet-octet splitting.

For the scalars we obtain in the SU(3) limit also that the corrections 
to the masses behave as
\begin{eqnarray}
\label{scal}
   &&(\Delta M^2_{a_0})_c=(\Delta M^2_{K_0^*})_c
     =(\Delta M^2_{88})_c,
     \nonumber \\
   &&(\Delta M^2_{08})_c=0, 
     \nonumber \\
   &&(\Delta M^2_{00})_c-(\Delta M^2_{88})_c\ne 0.
\end{eqnarray}  


\section*{\normalsize\it 3.3 Weak decay constants }

We use PCAC and the Gell-Mann-Oakes-Renner (GOR) \cite{Gell-Mann:1968} 
relation to extract the condensates. From PCAC the weak decay constants 
are given as 
\begin{equation}
\label{weak}
   f_\pi=\frac{M_u}{g}\ , \quad
   f_K=\frac{M_u+M_s}{2g}\ .
\end{equation}
Using this and the mass relations for $m_\pi$ and $m_K$, 
eqs.(\ref{mpi2}) and (\ref{mk2}), one obtains the GOR equations with 
some model corrections of higher order in the current quark masses 
and from which one identifies the condensates
\begin{eqnarray}
\label{GOR0}
   m^2_\pi f_\pi^2&\!\!\! =\!\!\!& m_u(h_u +2c_u)
       \left(1+\frac{m_u}{\Delta_u}\right)= 
       -2m_u \big <0|\bar{u} u|0\big >
       \left(1+\frac{m_u}{\Delta_u}\right), 
       \nonumber \\
   m^2_K f^2_K &\!\!\! =\!\!\!& 
   \frac{1}{4}(m_u +m_s)(h_u+2 c_u+h_s+2 c_s) 
   \left(1+\frac{m_u +m_s}{\Delta_u+\Delta_s}\right)
   \nonumber \\
   &\!\!\! =\!\!\!& -\frac{1}{2}(m_u +m_s) 
   \big <0|\bar{u} u + \bar{s} s|0\big > 
   \left(1+\frac{m_u +m_s}{\Delta_u+\Delta_s}\right).
\end{eqnarray}

Finally, by using the gap equations (\ref{gapfluc}) in ($\ref{GOR0}$) 
and expressing $I_0, I_1$ through $J_0(M_i^2), J_1(M_i^2)$ with 
(\ref{Iint}) we obtain the condensates as 
\begin{equation}
\label{conf}
   \big <0|\bar {q}_i q_i|0\big > = 
   -\frac{N_c}{4\pi^2}\ M_i J_0(M_i^2) +{\cal O}(J_2), 
\end{equation}
where the  ${\cal O}(J_2)$ terms are neglected, to conform with the 
truncation of the heat kernel series. We recall that the ${\cal
  O}(J_2)$ emerge from a property of the generalized heat kernel
series in which differences of $J_k(M_u^2)-J_k(M_s^2)$ are expressed
as an infinite series involving $J_{k+l}$, $l>0$ \cite{Osipov1:2001}.


\section*{\centerline{\large\sc 4. Numerical results and discussion}}

There are six parameters in the model, $m_u, m_s, G, \kappa, \Lambda, 
\lambda$. To see the effects of the new contribution, proportional to 
the cutoff $\lambda$, we compare pairwise in the sets (a,b), (c,d) and 
(e,f) of Tables 1-3 the results calculated with $\lambda=0$ and 
$\lambda\ne 0$, keeping the remaining input unchanged. In this way, 
within each pair of sets, a running value of $\lambda$ between the 
indicated ones, will interpolate smoothly between the calculated 
observables shown. In a-d we fix four parameters through
the pseudoscalar sector, $m_\pi, m_K, f_\pi, f_K$, and adjust
$\kappa$ through the quark condensate $\big <\bar u u\big >$. In sets 
(e,f) five parameters are fixed through $m_\pi, m_K, f_\pi, m_{\eta'}$, 
and the scalar $a_0$. Input is indicated through a *. 

\vspace{0.4cm}
\noindent
{\rm Table 1.}\\
{\footnotesize The main parameters of the model: current quark masses 
$m_u$, $m_s$, and corresponding constituent masses, $M_u$
and $M_s$ in MeV, couplings $G$ (in GeV$^{-2}$) and $\kappa$ 
(in GeV$^{-5}$), and two cutoffs $\Lambda, \lambda$ in GeV. The values
of condensates are given in MeV.}
\vspace{0.2cm}

\noindent
\begin{tabular}{lccccccccccc}
\hline 
& $m_u$ 
& $m_s$ 
& $M_u$
& $M_s$
& $G$  
& $-\kappa$ 
& $\Lambda$
& $\lambda$ 
& $-\big <\bar{u} u\big >^{1/3}$ 
& $-\big <\bar{s} s\big >^{1/3}$ \\ 
\hline
a&6.3 & 194 & 398 & 588 & 13.5 & 1300* & 0.82 & 0*   & 229 & 172 \\ 
b&6.3 & 194 & 398 & 588 & 13.5 & 1300* & 0.82 & 1.8* & 229 & 172 \\ 
c&6.3 & 194 & 398 & 588 & 13.4 & 1370* & 0.82 & 0*   & 229 & 172 \\ 
d&6.3 & 194 & 398 & 588 & 11.8 & 1370* & 0.82 & 1.7* & 229 & 172 \\
e&2.8 &  92 & 216 & 385 & 3.14 &  120  & 1.37 & 0*   & 302 & 314 \\ 
f&2.1 &  69 & 196 & 354 & 2.15 &   53  & 1.64 & 1.9* & 333 & 363 \\ 
\hline
\end{tabular}   
\vspace{0.8cm}

\vspace{0.4cm}
\noindent
{\rm Table 2.}\\
{\footnotesize The main characteristics of the light pseudoscalar 
mesons in MeV. The singlet-octet mixing angle $\theta_p$ is given in
degrees. }
\vspace{0.2cm}

\noindent
\begin{tabular}{lcccccccc}
\hline
& $ m_\pi     $
& $ m_K       $
& $ f_\pi     $
& $ f_K       $
& $ m_\eta    $
& $ m_{\eta'} $
& $\theta_p   $ \\ 
\hline
a & 138* & 494* & 92* & 114* & 476 &  986  & -14 \\ 
b & 138* & 494* & 92* & 114* & 487 &  958  & -15 \\ 
c & 138* & 494* & 92* & 114* & 480 & 1020  & -13 \\ 
d & 138* & 494* & 92* & 114* & 472 &  959  & -15 \\ 
e & 138* & 494* & 92* & 129  & 533 & 1097* & -1.2 \\ 
f & 138* & 494* & 92* & 129  & 540 & 1097* & 0.5 \\ 
\hline
\end{tabular}
\vspace{1cm}

\vspace{1.0cm}
\noindent
{\rm Table 3.}\\
{\footnotesize The characteristics of the light scalar nonet in MeV
and the singlet-octet mixing angle $\theta_s$ in degrees.}
\vspace{0.2cm}

\noindent
\begin{tabular}{lccccc}
\hline
& $ m_{a_0}\!\!\sim\! a_0(980)     $
& $ m_{K_0^*}\!\!\sim\! K_0^*(800) $
& $ m_{\sigma}\!\!\sim\! f_0(600)  $
& $ m_{\sigma'}\!\!\sim\! f_0(980) $
& $ \theta_s                 $ \\ 
\hline
a & 1040  & 1267 & 806 & 1438 & 24   \\ 
b &  981  & 1219 & 781 & 1427 & 24   \\ 
c & 1056  & 1280 & 805 & 1447 & 23.7 \\ 
d &  967  & 1208 & 762 & 1426 & 23.5 \\ 
e &  980* & 1029 & 413 & 1123 & 19.5 \\ 
f &  980* &  992 & 346 & 1073 & 18   \\ 
\hline
\end{tabular}
\vspace{1cm}

It is clear that the large differences observed among different pairs
of sets come from the leading order contribution. For example the 
condensates, $f_K$ and the $\eta$ mass are strongly dependent on the 
value of $\kappa$, which is one order of magnitude larger in the sets 
(a-d) as compared to set (e,f). This observation applies also to the 
scalar spectrum, with large changes resulting at leading order. They 
are best described in set (f), but this implies  rather large values 
for $f_K$ and the condensates. 

One observes however that the corrections, although small, have the 
correct trend, diminishing the splitting in the singlet-octet members 
of the pseudoscalar and scalar spectra. Comparing sets (a,b) with
(c,d), one sees that by enlarging the magnitude of $\kappa$, the
effect of the corrections get stronger, even for a smaller value of 
$\lambda$. 

We also remark the interesting fact that $\lambda$ cannot be
arbitrarily increased, maintaining the remaining input on observables 
fixed. At values quite close to the ones indicated, solutions cease to 
exist. There is an intrinsic constraint on the size of corrections. 

One might be struck by the large variation in the values of $\kappa$
and their relation to the convergence of the loop expansion 
(\ref{Zloop1}). This can be understood by identifying the
dimensionless expansion parameter of the series. Following
\cite{Osipov:2005a}, where standard methods are used to justify the 
stationary phase approach to functional integrals, one  obtains the 
dimensionless parameter 
\begin{equation}
\label{zeta}
   \zeta=\frac{\kappa^2}{32G^3}\left( \frac{\lambda}{2\pi}\right)^4.
\end{equation}
Here the group structure factor $3/16$ of $\Phi_{abc}$ in 
eq.(\ref{Phi}) as well as the factor $1/3!$ appearing at each 
order in $(\ref{Zloop1})$ have been taken into account. Note also that 
each term of the expansion carries a further suppression factor $1/n!$
In the present case $n=2$ so that one get for sets (b), (d), (f) that 
$\zeta^2/2$ is $0.0096,\ 0.021,\ 0.0023$ respectively. This 
attests for a fast convergence of the series. 

For a comparison with empirical values, we take from \cite{RPP:2004}: 
\begin{eqnarray}
   && m_u=1.5 \div 4\ \mbox{MeV}, \nonumber \\ 
   && m_d=4 \div 8\ \mbox{MeV},  \nonumber \\ 
   && m_s=80 \div 130\ \mbox{MeV}, \nonumber \\
   && m_{\pi^\pm}=139.57018\pm 0.00035\ \mbox{MeV}, \nonumber \\  
   && m_{K^\pm}=493.677\pm 0.016\ \mbox{MeV}, \nonumber \\ 
   && m_\eta=547\pm 0.12\ \mbox{MeV}, \nonumber \\
   && m_{\eta'}=957.78\pm 0.14\ \mbox{MeV},  
\end{eqnarray}
for the masses in the low lying pseudoscalar sector. The weak decay 
constants
$f^{e}_\pi = 130.7\pm 0.1\pm 0.36\ \mbox{MeV}$, 
$f^{e}_K = 159.8\pm 1.4\pm 0.44\ \mbox{MeV}$ 
relate to ours through a $\sqrt{2}$ normalization factor, thus 
$f_\pi \simeq 92.4\ \mbox{MeV}$ and 
$f_K \simeq 113\ \mbox{MeV}$.

The low-lying scalar masses are presently\footnote{As several data 
sets are presented in \cite{RPP:2004}, please consult it for details. 
Here we indicate the lowest and the highest values collected from all 
samples.}:
\begin{eqnarray}   
   && m_{a_0(980)} = 984.7\pm 1.2\ \mbox{MeV}, \nonumber \\  
   && m_{f_0(600)} = 400 \div 1200\ \mbox{MeV}, \nonumber \\ 
   && m_{f_0(980)} = 980 \pm 10\ \mbox{MeV}, \nonumber \\ 
   && m_{K_0^*(800)} = 701 \div 970\ \mbox{MeV}.
\end{eqnarray} 

The recent update of the light-quark condensate is 
$\big< (\bar{u}u+\bar{d}d)/2 \big > (1\ \mbox{GeV})=-(242\pm 15\ 
\mbox{MeV})^3$, the flavour breaking ratio is known to be 
$\big < \bar ss\big>/\big< (\bar{u}u+\bar{d}d)/2 \big > =
0.8 \pm 0.3$ \cite{Jamin:2002}.

Finally, as our numerical calculations do not differ significantly 
from the leading order values, we refer to \cite{Osipov:2004NPA} where 
a thorough discussion of our leading order results has been done in 
comparison with the ones obtained from other approaches 
\cite{Tornqvist:1999}-\cite{Schechter:2002}.


\section*{\centerline{\large\sc 5. Conclusions}}

The bosonization of the model combining the NJL and the 't Hooft 
multi-quark interactions leads to corrections associated with the 
stationary phase integration over auxiliary bosonic variables in the 
functional integral of the theory. The purpose of the present work has 
been to quantify NLO corrections, and to study their phenomenological 
effect on the mass spectrum of light pseudoscalar and scalar mesons.  
 
To this end the first correction to the tree level effective action
has been considered. We have obtained the linear and quadratic terms
(in the external mesonic fields) of the NLO Lagrangian. The group 
structure of the  $SU(2)\times U(1)$ flavour symmetry considered leads
to quite intricate expressions for the mass corrections. We have
shown in a transparent way that they comply with the QCD low energy 
theorems. We have calculated the mass spectra of the low lying 
pseudoscalars and scalars, quark condensates and weak decay constants 
$f_\pi, f_K$. The corrections are small and improve slightly the
leading order results. 

We conclude from these calculations that the series considered is well 
convergent. It is an important conclusion, because it justifies
the leading order estimates made before, on one hand, and reports
on the selfconsistency of the stationary phase approach applied to the
bosonization of effective multi-quark interactions, on the
other hand.
  
At the same time one may still expect some noticeable effects of the 
NLO terms which can show themselves in three and higher order mesonic 
amplitudes, especially in the cases where there is a strong
cancellation between the tree level contributions.


\section*{\centerline {\large\sc Acknowledgements}}

This work has been supported by grants provided by Funda\c c\~ao para
a Ci\^encia e a Tecnologia, POCTI/FNU/50336/2003, POCI/FP/63412/2005. 
This research is part of the EU integrated infrastructure initiative 
Hadron Physics project under contract No.RII3-CT-2004-506078. A. A. 
Osipov also gratefully acknowledges the Funda\ca o Calouste Gulbenkian 
for financial support.


\newpage
\noindent
{\bf Appendix}
\vspace{0.5cm}

The equations and algebra leading to the coefficients $h_{ab...}^{(i)}$ 
of the series for $s_a$ and $p_a$ can be found in \cite{Osipov:2004}. 
The explicit expressions for the case of one and two lower indices are 
also given there. In this appendix we collect explicit expressions for 
the coefficients $h_{abc}^{(1)}$ and $h_{abc}^{(2)}$ entering in the 
expansion of $s_a$ and needed for the evaluation of meson mass terms. 
Due to the trace structure of (\ref{M2}) and since ${\bar L}_0$ 
contributes only in the blockdiagonal, with nonvanishing entries in the 
diagonal and elements of $0,8$ mixing , only the elements of the diagonal 
and $(0,8)$ mixing of $L_2$ will contribute to the mass terms. For those 
we need:
\begin{eqnarray}
   && h_{0aa}^{(1)}=\frac{\kappa}{16 \sqrt{6} G^3}\ 
      \frac{1+\omega_s-2 \omega_u}{\mu_+ (1-\omega_s)^2} 
      \qquad \mbox{for  } a\in 1,2,3. \nonumber \\
   && h_{0aa}^{(1)}=\frac{\kappa}{16 \sqrt{6} G^3}\ 
      \frac{1}{\mu_+ (1-\omega_u)} \qquad
      \mbox{for  }  a\in 4,5,6,7. \nonumber \\
   && h_{8aa}^{(1)}=-\frac{\kappa}{16 \sqrt{3} G^3}\ 
      \frac{1+\omega_s+ \omega_u}{\mu_+ (1-\omega_s)^2}
      \qquad \mbox{for  }  a\in 1,2,3. \nonumber \\ 
   && h_{8aa}^{(1)}=\frac{\kappa}{32 \sqrt{3} G^3}\ 
      \frac{1+2 \omega_u}{\mu_+ (1-\omega_u)^2} 
      \qquad \mbox{for  }  a\in 4,5,6,7. 
\end{eqnarray}
\begin{eqnarray}
   && h_{000}^{(1)}=-\frac{\kappa}{8 \sqrt{6} G^3 
      \mu_+^3}\ (1+\omega_s-2 \omega_u)(1-\omega_u)^2
      \nonumber \\
   && h_{088}^{(1)}=\frac{\kappa}{16 \sqrt{6} G^3 
      \mu_+^3}\ (1+2 \omega_u)[1+\omega_s(1-2 \omega_u)]
      \nonumber \\
   && h_{008}^{(1)}=h_{080}^{(1)}=h_{800}^{(1)}=
      \frac{\kappa}{8 \sqrt{3} G^3 \mu_+^3}\ \omega_u 
      (\omega_u- \omega_s)(1-\omega_u) \nonumber \\
   && h_{808}^{(1)}=h_{880}^{(1)}=\frac{\kappa}{16 \sqrt{6} 
      G^3 \mu_+^3}\ (1+ \omega_s+2\omega_u-4 \omega_s\omega_u^2)
      \nonumber \\
   && h_{888}^{(1)}=\frac{\kappa}{16 \sqrt{3} G^3 \mu_+^3}\ 
      (1+\omega_u+ \omega_s)(1+2\omega_u)^2
\end{eqnarray}
\begin{eqnarray}
   && h_{0aa}^{(2)}=-\frac{\kappa}{16 \sqrt{6} G^3}\ 
      \frac{1+\omega_s-2\omega_u}{\mu_+ (1+\omega_s)^2}
      \qquad \mbox{for  }  a\in 1,2,3. \nonumber \\
   && h_{0aa}^{(2)}=-\frac{\kappa}{16 \sqrt{6} G^3}\ 
      \frac{1- \omega_u}{\mu_+ (1+\omega_u)^2} \qquad
      \mbox{for  }  a\in 4,5,6,7. \nonumber \\
   && h_{8aa}^{(2)}=\frac{\kappa}{16 \sqrt{3} G^3}\ 
      \frac{1+\omega_s+ \omega_u}{\mu_+ (1+\omega_s)^2}
      \qquad \mbox{for  }  a\in 1,2,3. \nonumber \\ 
   && h_{8aa}^{(2)}=-\frac{\kappa}{32 \sqrt{3} G^3}\ 
      \frac{1+2 \omega_u}{\mu_+ (1+\omega_u)^2} 
      \qquad \mbox{for  }  a\in 4,5,6,7.
\end{eqnarray}

\begin{eqnarray}
   && h_{000}^{(2)}=\frac{\kappa}{24 \sqrt{6} G^3 \mu_+\mu_-^2}\ 
      (1+ \omega_u)(3+\omega_u-\omega_s+3\omega_u\omega_s-6\omega_u^2)
      \nonumber \\
   && h_{088}^{(2)}=-\frac{\kappa}{48 \sqrt{6} G^3 \mu_+\mu_-^2}\ 
      (1-2 \omega_u)[3-4\omega_u -\omega_s(5-6 \omega_u)]
      \nonumber \\
   && h_{008}^{(2)}=h_{080}^{(2)}=-\frac{\kappa}{24 \sqrt{3} 
      G^3 \mu_+\mu_-^2}\ (\omega_u- \omega_s)(1+\omega_u-3 \omega_u^2)
      \nonumber \\
   && h_{800}^{(2)}=\frac{\kappa}{24 \sqrt{3} G^3 \mu_+\mu_-^2}\
      (\omega_u- \omega_s)(1+\omega_u)(2+ 3\omega_u)
      \nonumber \\
   && h_{808}^{(2)}=h_{880}^{(2)}=-\frac{\kappa}{48 \sqrt{6} G^3 
      \mu_+\mu_-^2}\ (3+2 \omega_u-4\omega_u^2+\omega_s 
      (1-8\omega_u-12\omega_u^2)) \nonumber \\
   && h_{888}^{(2)}=-\frac{\kappa}{48 \sqrt{3} G^3 \mu_+\mu_-^2}\ 
      (1-2\omega_u)(3+\omega_u-\omega_s-6\omega_u\omega_s-6\omega_u^2)
      \nonumber \\
\end{eqnarray}
where 
\begin{equation}
   \omega_i=\frac{\kappa h_i}{16 G}, \qquad
   \mu_{\pm}=1\pm\omega_s-2\omega_u^2.
\end{equation}


\end{document}